\documentclass[A4paper,12pt,conference]{report}

\usepackage[page,toc]{appendix}
\usepackage[a4paper,portrait,bottom=2.5cm,top=2.5cm,left=3.8cm,right=2.5cm]{geometry}

\usepackage{mystyle}
\usepackage{pythonhighlight}
\usepackage{balance}
\usepackage{cite}
\usepackage{float}
\usepackage{graphicx}
\usepackage{tikz}
\usetikzlibrary{shapes, arrows.meta, positioning, backgrounds, fit}
\usetikzlibrary{calc} 
\usepackage{booktabs} 
\usepackage{algorithm}
\usepackage{algorithmic}
\usepackage{multirow}
\usepackage{gensymb} 
\usepackage{color}
\usepackage{enumerate}
\usepackage{makecell}

\usepackage{dutchcal}

\DeclareMathAlphabet\mathbfcal{OMS}{cmsy}{b}{n}

\usepackage{setspace}
\begin{document}
\onehalfspacing
\begin{center}
		{\fontsize{14}{16.8}
		\bf Multi-modal Deep Learning}
	\end{center}

	\vspace{3mm}
	\begin{center}
		by
	\end{center}
	
	\vspace{3mm}
	\begin{center}
		\bf CHEN YUHUA \\ \quad\\
		A0259988W
	\end{center}

	\vspace{4mm}
	\begin{center}
		MSc Report for EE5003
	\end{center}

	\vspace{3mm}
	\begin{center}
		\bf Master of Electrical Engineering
	\end{center}

	\vfill
	\begin{center}
		\bf 2023 \\March 30th
	\end{center}
	
	\begin{figure}[htbp!]
		\centering
		\includegraphics[width=.8\textwidth]{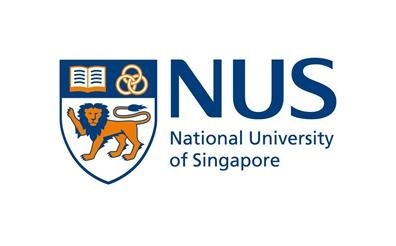} \\
	\end{figure}
			
	\begin{center}
		College of Design and Engineering
	\end{center}
	
	\begin{center}
		National University of Singapore
	\end{center}
    \thispagestyle{empty}


%
%
%

%
%

\markboth{Journal of \LaTeX\ Class Files,~Vol.~11, No.~4, December~2012}%
{Shell \MakeLowercase{\textit{et al.}}: Bare Demo of IEEEtran.cls for Journals}
%




\begin{abstract}
This article investigates deep learning methodologies for single-modality clinical data analysis, as a crucial precursor to multi-modal medical research. Building on Guo JingYuan's work, the study refines clinical data processing through Compact Convolutional Transformer (CCT), Patch Up, and the innovative CamCenterLoss technique, establishing a foundation for future multimodal investigations.

The proposed methodology demonstrates improved prediction accuracy and attentiveness to critically ill patients compared to Guo JingYuan's ResNet and StageNet approaches.
Novelty that using image-pretrained vision transformer backbone to perform transfer learning time-series clinical data.The study highlights the potential of CCT, Patch Up, and novel CamCenterLoss in processing single-modality clinical data within deep learning frameworks, paving the way for future multimodal medical research and promoting precision and personalized healthcare.
\end{abstract}

\addtocontents{toc}{\protect\thispagestyle{empty}}

\tableofcontents
\thispagestyle{empty} 

\chapter{Introduction}

\setcounter{page}{1}
\section{ICH mortality prediction}
	\indent Intracerebral hemorrhage (ICH) is a prevalent neurological emergency, accounting for 6.5$\%$ - 19.6$\%$  of all strokes and exhibiting higher morbidity and mortality rates compared to ischemic strokes \cite{feigin2003stroke}. Each year, around 2 million people are affected by ICH \cite{cordonnier2018intracerebral,van2010incidence}. With a high mortality rate, approximately 35$\%$ of ICH patients die within 7 days and 50$\%$ within 30 days \cite{van2010incidence,carolei1997high}. Early mortality prediction or warning for ICH patients is crucial for assessing their condition and evaluating novel treatments, interventions, and healthcare policies.

    \indent Conventional scores, such as the ICH score and Acute Physiology and Chronic Health Evaluation (APACHE) II system, have been developed to predict in-hospital mortality, relying on statistical analysis to identify relevant covariates from preselected features by domain experts \cite{monteiro2018using}. These methods present drawbacks, such as requiring domain expertise, rendering them less accessible to nonprofessional users. Moreover, the calculations can be tedious, leading to simplification in practice and subsequent performance deterioration.
    
    \indent In contrast, machine learning methods alleviate manual calculation burdens and enable in-depth analysis using big data. These approaches can uncover hidden information based on learned patterns, enhancing our understanding of the disease. The current study \cite{nie2021mortality} compares conventional scores and basic machine learning techniques like random forest, nearest neighbors, and adaboost. Although these methods are relatively simple, deep learning has proven to be more powerful.
    
    \indent Building based the work of former master's student Guo JingYuan, I have further improved the deep learning model for ICH patients' clinical data analysis. This refined model outperforms Guo JingYuan's model and demonstrates enhanced robustness due to the introduction of noise to input values during training. In this report, we present our improved deep learning model, detailing its development, implementation, and the advantages it offers in predicting ICH patients' mortality and informing healthcare decisions.

\section{Transformer Literature Review}
    \indent The transformer architecture has emerged as a powerful tool across various fields and applications, ranging from natural language processing to computer vision and beyond. This literature review aims to provide an overview of recent advancements and applications of transformer models in diverse areas.

    \indent Ćalasan et al. \cite{calasan2019} evaluate the use of Chaotic Optimization Approach on both nameplate and load data, contrasting with literature approaches that rely on different estimation techniques and data sources. Strain et al. \cite{strain2019} explore the analysis of ZVS in a GaN-based LLC resonant converter with two series-parallel connected transformers. Palomino and Parvania \cite{palomino2020} propose a data-driven risk assessment method for quantifying transformer and secondary conductor overload conditions due to high levels of EV charging demand and rooftop solar PV.

    \indent Li et al. \cite{li2021} present an empirical comparison of RNN-T, RNN-AED, and Transformer-AED models in both non-streaming and streaming modes. Park et al. \cite{park2021} address the suboptimal feature embedding issue in existing vision transformers by proposing a novel vision Transformer that utilizes a low-level CXR feature corpus for extracting abnormal CXR features. Gao et al. \cite{gao2021} introduce an explainable deep learning network for classifying COVID from non-COVID based on 3D CT lung images.

    \indent Pan et al. \cite{pan2021} examine a multi-domain integrative swin transformer network for sparse-view tomographic reconstruction, resulting in the development of a Multi-domain Integrative Swin Transformer network (MIST-net) to improve image quality from sparse-view data. Xie et al. \cite{xie2022} investigate deep learning-based multi-user semantic communication systems for transmitting single-modal and multimodal data. Lin et al. \cite{lin2022} propose a transformer fault diagnosis method based on the IFCM-DNN adjudication network, demonstrating its effectiveness in improving diagnostic accuracy under unbalanced data sets. Keitoue et al. \cite{keitoue2018} present an online transient overvoltage monitoring system (TOMS) for power transformers, capable of continuously recording real-time transient overvoltages.

    \indent In conclusion, the transformer architecture has demonstrated remarkable versatility and efficacy in addressing various challenges across different domains. This literature review highlights the ongoing innovation and potential of transformer models in solving complex problems, as well as their ability to adapt and improve upon existing methodologies.

\chapter{Related work}

\section{MIMIC III Benchmark}
    The lack of accessible public benchmarks presents challenges when comparing the performance of different models, particularly in healthcare contexts. Despite the steady development of machine learning research in this field, a comprehensive public benchmark remains scarce. To address this issue, the MIMIC-III benchmark from \cite{harutyunyan2019multitask} provides a standardized data cleaning and extraction process for the MIMIC-III database. 

\section{Machine Learning Methods}
    Numerous machine learning approaches have been employed to analyze ICH patients from the MIMIC-III database in order to evaluate their predictive capabilities regarding in-hospital mortality. In the study by \cite{nie2021mortality}, a variety of machine learning techniques, including random forest, adaboost, nearest neighbors, and neural networks, were investigated. The results indicated that all machine learning models outperformed the conventional clinical method, APACHE II score, with random forest exhibiting the best performance among them. This evidence suggests that machine learning techniques offer superior assistance to clinicians in assessing patients' health conditions compared to traditional clinical scoring systems.
    
\section{Stagenet Baseline}
    In the study on StageNet, the authors directly utilize the raw data as input to the LSTM cell, which is then processed through a stage-adaptive convolution to obtain mortality prediction.\cite{gao2020stagenet} Jingyuan's work builds upon the foundation of StageNet by incorporating a ResNet backbone as a feature extractor to further enhance the performance of the network.\cite{Guo2022} This improvement aims to leverage the powerful representation learning capabilities of ResNet, allowing for more effective feature extraction and ultimately leading to more accurate mortality predictions.

\begin{figure}[htbp!]
        \centering
        \includegraphics[width=1\textwidth]{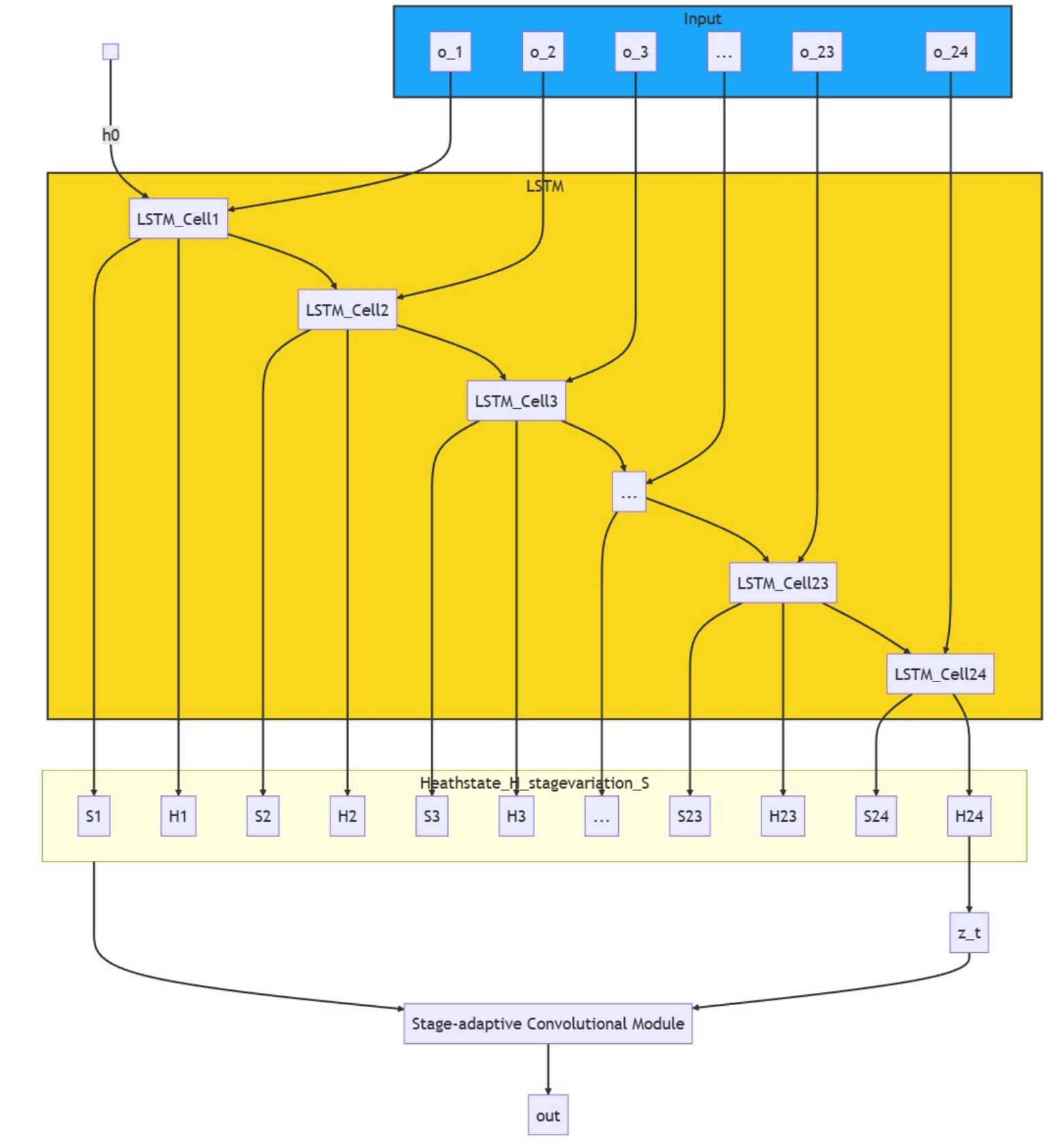}
        \caption{ stage-net origin structure \cite{gao2020stagenet}}
        \label{fig 2.1:stagenet}
    \end{figure}

    \indent The basic structure of original stage-net \cite{gao2020stagenet}  is shown in Figure \ref{fig 2.1:stagenet} ,including input layer, hidden layers and output layer.
    
\begin{figure}[htbp!]
        \centering
        \includegraphics[width=1\textwidth]{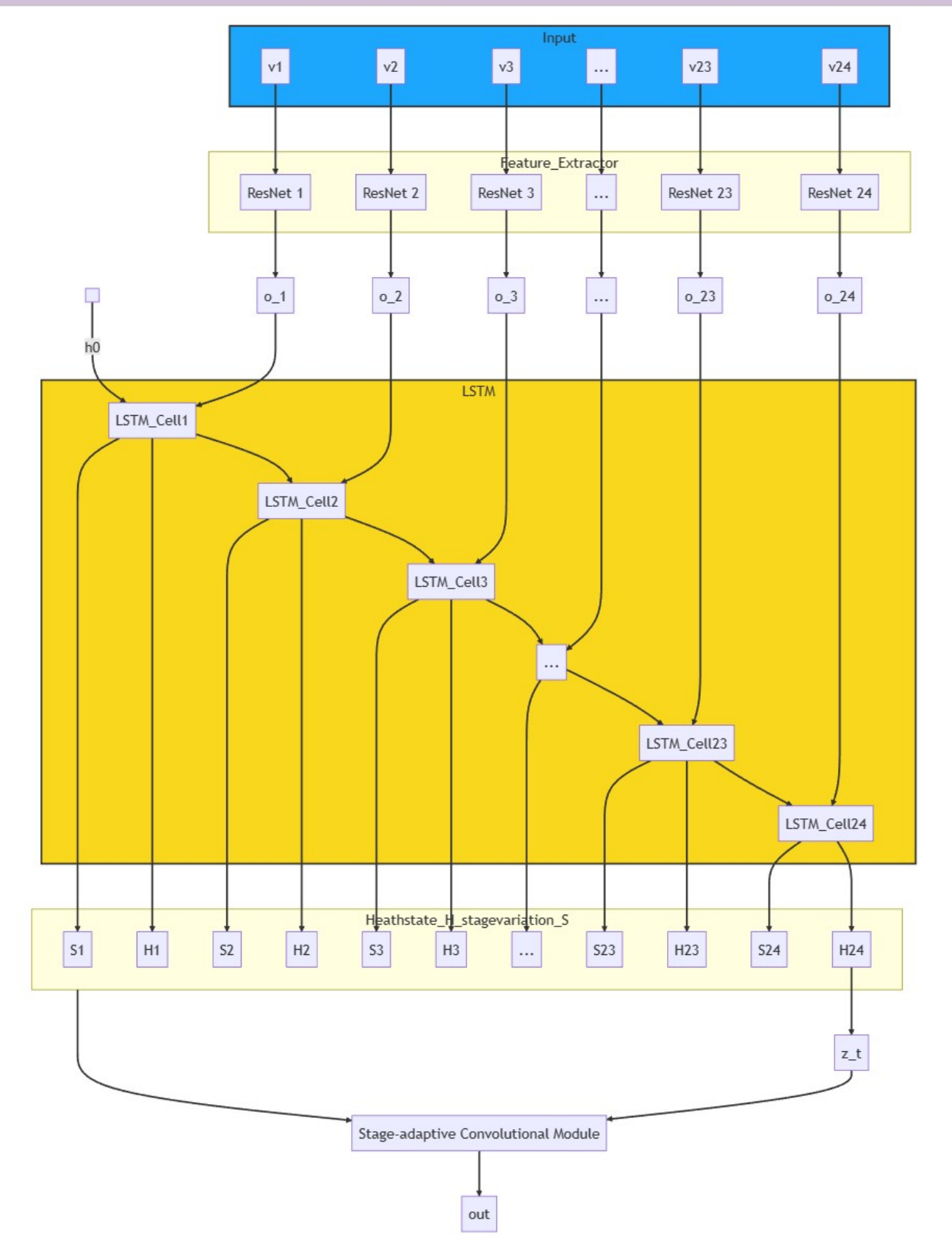}
        \caption{ Baseline JingYuan's structure \cite{Guo2022}}
        \label{fig 2.2:JingYuan's structure}
    \end{figure}

    \indent The baseline structure of this report, which is JingYuan's structure \cite{Guo2022} shown in Figure \ref{fig 2.2:JingYuan's structure}
\chapter{Proposed method}
    \section{Model Structure Overview}
		
\begin{figure}[htbp!]
        \centering
        \begin{tikzpicture}[  node distance=1.5cm,  data/.style={rectangle, draw, minimum width=1cm, minimum height=1cm},  cell/.style={ellipse, draw=blue!50!black, fill=blue!30!white, opacity=1, minimum width=1.5cm, minimum height=1cm},  arrow/.style={->, thick}]

\node[data] (v1) {$v_{1}$};
\node[data, right=of v1] (v2) {$v_{2}$};
\node[data, right=of v2] (v3) {$v_{3}$};
\node[right=of v3] (dots) {$\cdots$};
\node[data, right=of dots] (v23) {$v_{23}$};
\node[data, right=of v23] (v24) {$v_{24}$};

\node[cell, above=of v1] (cell1) {ResNet 1};
\node[cell, above=of v2] (cell2) {ResNet 2};
\node[cell, above=of v3] (cell3) {ResNet 3};
\node[cell, above=of v23] (cell23) {ResNet 23};
\node[cell, above=of v24] (cell24) {ResNet 24};
\node[] (cell_dots) at ($(cell3)!0.5!(cell23)$) {$\cdots$};

\draw[arrow] (v1) -- (cell1);
\draw[arrow] (v2) -- (cell2);
\draw[arrow] (v3) -- (cell3);
\draw[arrow] (v23) -- (cell23);
\draw[arrow] (v24) -- (cell24);

\node[data, above=of cell1] (o1) {$o_{1}$};
\node[data, above=of cell2] (o2) {$o_{2}$};
\node[data, above=of cell3] (o3) {$o_{3}$};
\node[data, above=of cell23] (o23) {$o_{23}$};
\node[data, above=of cell24] (o24) {$o_{24}$};
\node[right=of o3] (odots) {$\cdots$};

\node[data, above=of o1] (concat1) {};
\node[data, above=of o2] (concat2) {};
\node[data, above=of o3] (concat3) {};
\node[data, above=of o23] (concat23) {};
\node[data, above=of o24] (concat24) {};

\begin{scope}
\node[ draw, inner sep=10pt,rounded corners,draw=orange!50!black, fill=orange!30!white,fit=(concat1) (concat24), opacity=1, font=\fontsize{18}{16}\selectfont] (concat) {Data Reconstruction};
\end{scope}

\begin{scope}[shift={(4,10)}]
  \filldraw[fill=red!30!white, draw=red!50!black] (0,0) rectangle (3,3);
  \filldraw[fill=green!30!white, draw=green!50!black] (0.5,0.5) rectangle (3.5,3.5);
  \filldraw[fill=blue!30!white, draw=blue!50!black] (1,1) rectangle (4,4);
  \node [above= of concat] (output){};
  \node[draw=none, blue, font=\fontsize{18}{16}\selectfont] at (2.5,2.5)(map) {\color{white}$224 \times 224$};
\end{scope}
\begin{scope}
\node[ draw, inner sep=10pt,rounded corners,draw=orange!50, fill=orange!30,above=of map, opacity=1,font=\fontsize{18}{16}\selectfont] (cct) {Imagenet Pretrained Compact Convolutional Transformer};
\end{scope}

\draw[arrow] (cell1) -- (o1);
\draw[arrow] (cell2) -- (o2);
\draw[arrow] (cell3) -- (o3);
\draw[arrow] (cell23) -- (o23);
\draw[arrow] (cell24) -- (o24);

\draw[arrow] (o1) -- (concat1);
\draw[arrow] (o2) -- (concat2);
\draw[arrow] (o3) -- (concat3);
\draw[arrow] (o23) -- (concat23);
\draw[arrow] (o24) -- (concat24);
\draw[arrow] (concat) -- (output);
\draw[arrow] (map) -- (cct);
\node[data, above=of cct] (p1) at (concat1 |- cct.north) {$p_{1}$};
\node[data, above=of cct] (p2) at (concat2 |- cct.north) {$p_{2}$};
\node[data, above=of cct] (p3) at (concat3 |- cct.north) {$p_{3}$};
\node[right=of p3] (pdots) {$\cdots$};
\node[data, above=of cct] (p23) at (concat23 |- cct.north) {$p_{23}$};
\node[data, above=of cct] (p24) at (concat24 |- cct.north) {$p_{24}$};


\begin{scope}[on background layer]
  \node[rectangle, draw=yellow!50!black, rounded corners, fill=yellow!20!white, opacity=1, fit=(cell1) (cell24), inner sep=10pt, label=below:Feature Extractor] {};
\end{scope}
\begin{scope}[on background layer]
  \node[rectangle, draw=blue!50!black, rounded corners, fill=blue!20!white, opacity=0.5, fit=(p1) (p24), inner sep=10pt, label=below:pseudo sequence](seq) {};
\end{scope}
\draw[arrow] (cct) |- ([yshift=1cm]seq.south) ;
\begin{scope}
  \node[rectangle, draw=green!50!black, rounded corners, fill=green!20!white, opacity=1, above=of seq, opacity=1,font=\fontsize{18}{16}\selectfont,inner sep=10pt](conv) {Stage-adaptive Convolutional Module};
\end{scope}
\draw[arrow] (seq) -- (conv);

\node[data, above=of p24] (zt) {$z_{t}$};
\draw[arrow] (p24) -- (zt);
\draw[arrow] (zt) |- (conv.east);

\begin{scope}
  \node[rectangle, draw=red!50!black, rounded corners, fill=red!20!white, opacity=1, above=of conv, opacity=1,font=\fontsize{18}{16}\selectfont,inner sep=10pt](out) {Mortalytiprediction(Output)};

\end{scope}
\draw[arrow] (conv) |- (out.south);

\end{tikzpicture}
        \caption{The model architecture.}
        \label{fig3.1:stage_tran_net}
    \end{figure}
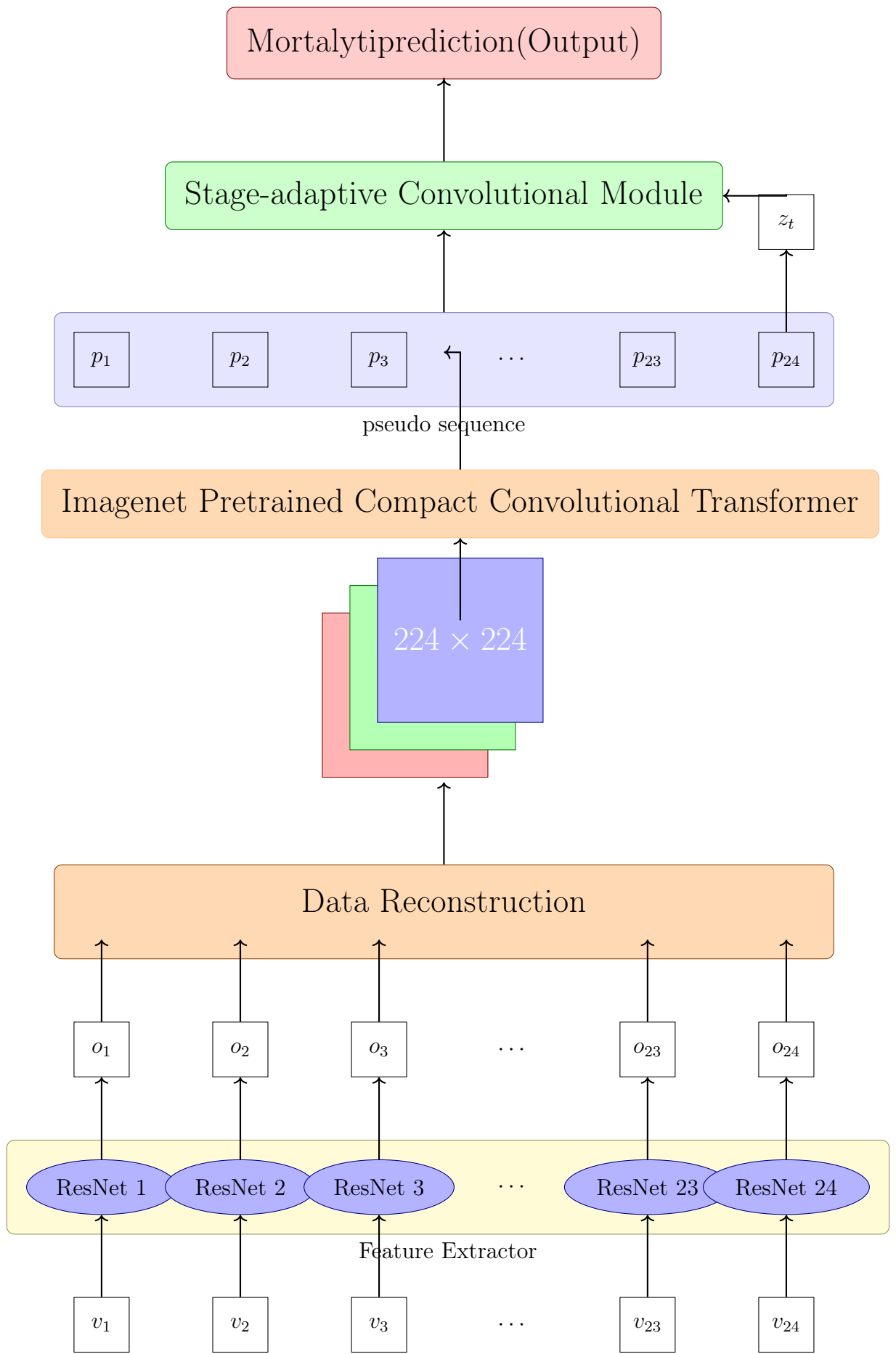

    \noindent As illustrated in Figure \ref{fig3.1:stage_tran_net}, the 24-hour, 812-variable dataset for each patient (referred to as a sample) is divided into 24 separate inputs, ranging from $\mathbf{v_1}$ to $\mathbf{v_{24}}$. These inputs are then processed through the ResNet backbone (specifically, ResNet-18) to serve as a feature extractor. Consequently, 24 extracted feature maps are generated, denoted as $\mathbf{o_1}$ to $\mathbf{o_{24}}$. The fully connected layer is utilized to reconstruct these maps, resulting in a (3, 224, 224) feature map for each patient.
    Here are some novelty that using image-pretrained vision transformer backbone to perform transfer learning time-series clinical data.

    Employing the generated feature map, it is subsequently fed into the ImageNet Pretrained Compact Convolutional Transformer (CCT) \cite{hassani2021cct}, which yields a 7,200-dimensional feature vector. This vector is then reshaped to create a (24, 300) pseudo-sequential feature map, with components ranging from $\mathbf{p_1}$ to $\mathbf{p_{24}}$. Finally, the Stage Adaptive Convolutional Module \cite{gao2020stagenet} is applied to obtain the predicted mortality outcome.

    \begin{table}[ht]
    \centering
    \caption{Notation Definition Table}
    \label{tab:notation_definition}
    \begin{tabular}{ll}
    \toprule
    Symbol & Definition \\
    \midrule
    $v_i$ & The $i$-th hour input variables of patient \\
    $o_i$ & The $i$-th output of Feature extractor \\
    $p_i$ & The $i$-th pseudo sequence feature for Stage-adaptive Convolutional Module \\
    $z_t$ & Last input of pseudo sequence \\
    \bottomrule
    \end{tabular}
    \end{table}

    \section{Compact Convolutional Transformer}
    In this report, we delve into the Compact Convolutional Transformer (CCT) model, as detailed in Hassani et al. \cite{hassani2021cct}. Specifically, we focus on the CCT-14/7x2 variant, which comprises 14 Transformer encoder layers and a 2-layer convolutional tokenizer with a 7×7 kernel size. It is worth noting that the model employed is an ImageNet-1k pre-trained checkpoint provided by Hassani et al. \cite{hassani2021cct}. For a concise illustration, we refer to their figure depicted below in Figure \ref{fig 3.2:CCT}. The main distinction between the original ViT \cite{dosovitskiy2020image} and CVT lies in the removal of the class token in the latter. Furthermore, CCT replaces the patching process with a convolutional process as a tokenizer, compared to CVT.

    \begin{figure}[!htbp]
    \centering
    \includegraphics[width=1\textwidth]{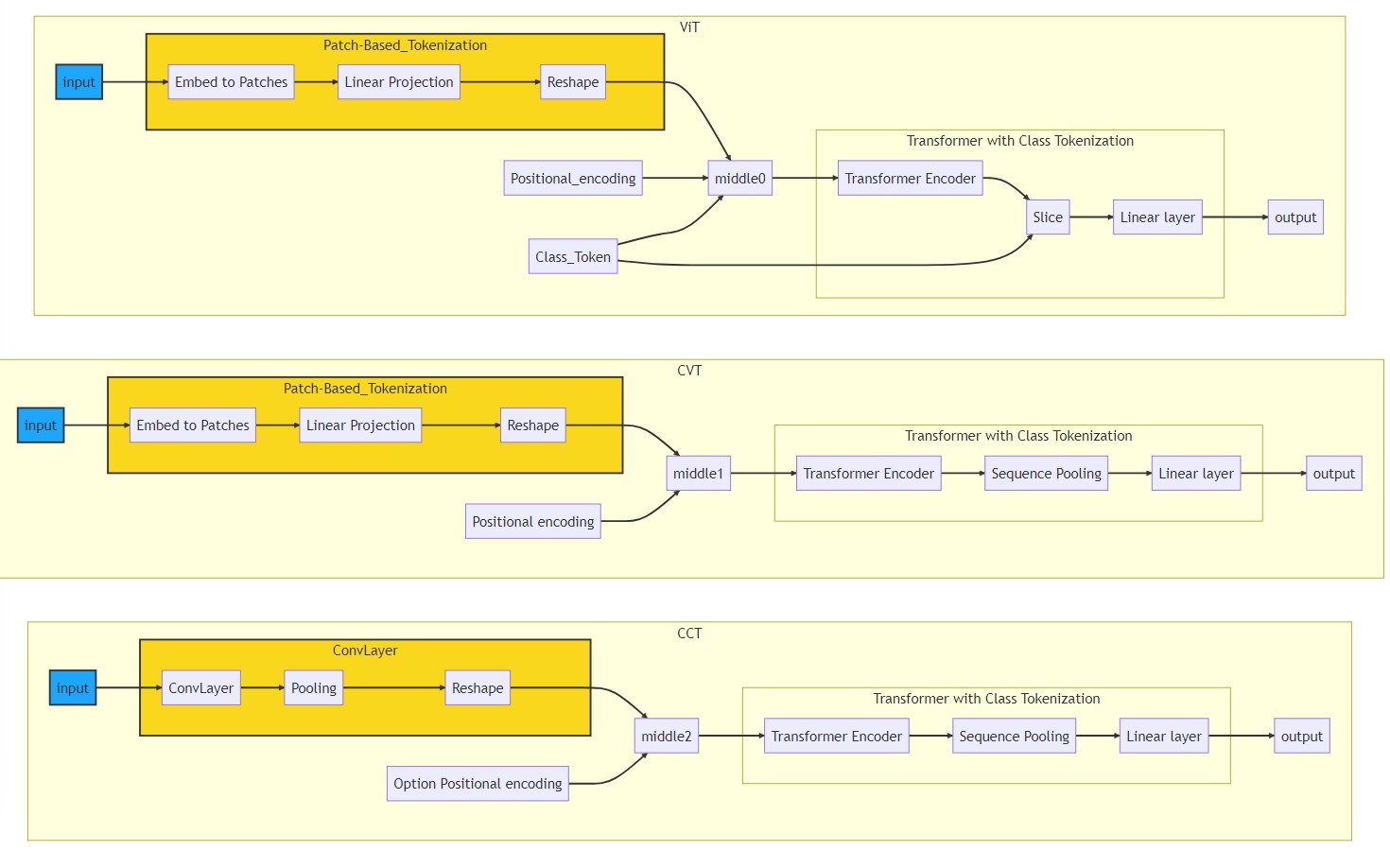}
    \caption{Comparison of CCT to CVT \cite{hassani2021cct}}
    \label{fig 3.2:CCT}
    \end{figure}
    
    \subsection{Convolutional Tokenizer}
    To imbue the model with an inductive bias, we replace the patch and embedding components in ViT-Lite and CVT with a rudimentary convolutional block. This block adheres to a traditional design, incorporating a single convolution, ReLU activation, and max-pooling operation. This approach enables the model to effectively retain local spatial information \cite{hassani2021cct}. Additionally, the implementation of this convolutional block enhances the model's flexibility compared to patching, potentially yielding superior performance on non-image data such as feature maps in hidden layers. Given an image or feature map \(x \in \mathbb{R}^{H \times W \times C}\), perform the operation as \eqref{eq:tokenization}
    
    \begin{equation}
    x_0 = \text{MaxPool}(\text{ReLU}(\text{Conv2d}(x)))
    \label{eq:tokenization}
    \end{equation}
    
    In our case, input should be \(x \in \mathbb{R}^{224 \times 224 \times 3}\), and after applying Equation \ref{eq:tokenization} twice followed by a flattening permutation, we obtain a positional embedded feature of shape \(o \in \mathbb{R}^{196 \times 384}\). It is important to note that we freeze the parameters of the pre-trained CCT tokenizer during training, with justification provided in the Experiment section. Here, 196 represents the position, and 384 denotes the feature dimension.
    
    \subsection{Convolutional Classifier}
    In our specific case, the convolutional classifier is an ordered concatenation of 14 Transformer encoder layers followed by a fully connected layer.
    \section{Patch Up}
    In this article\cite{faramarzi2020patchup}, PatchUp, a block-level feature space regularization technique for convolutional neural networks (CNNs), is introduced to address the high generalization gap encountered in deep learning models using limited labeled training data. PatchUp enhances the robustness of CNN models by selecting contiguous blocks from feature maps of randomly chosen sample pairs. Experimental results demonstrate that PatchUp outperforms or matches state-of-the-art regularizers across multiple datasets, improving generalization capabilities in image classification, deformed image classification, and adversarial attack resistance.
    \begin{figure}[!htbp]
    \centering
    \includegraphics[width=1\textwidth]{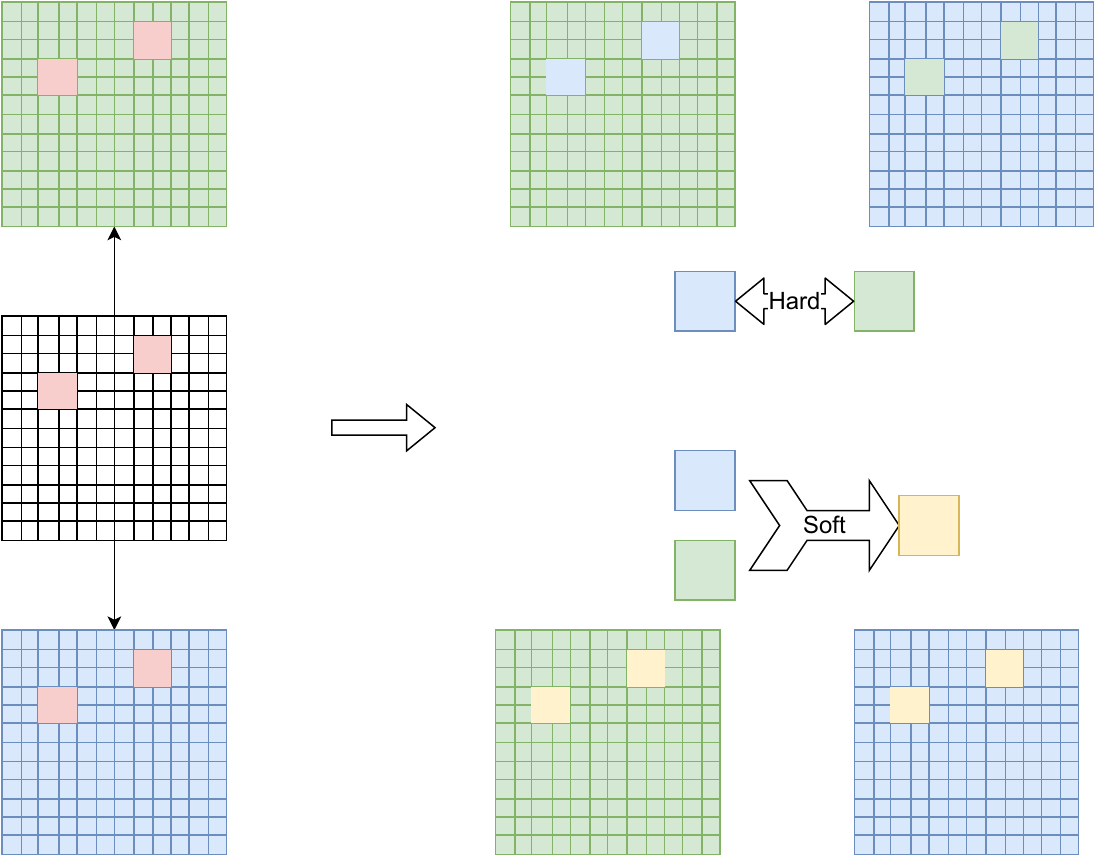}
    \caption{PatchUp process for two hidden representations associated with two samples  \cite{faramarzi2020patchup}}
    \label{fig 3.3:Patch up}
    \end{figure}
    PatchUp incorporates two distinct operation modes: Hard PatchUp and Soft PatchUp. Hard PatchUp operates by swapping contiguous blocks in feature maps of randomly selected sample pairs, while Soft PatchUp functions by blending these chosen contiguous blocks. Specifically, Hard PatchUp's operation at layer k is defined as in \eqref{eq:Hard patch up}:
    \begin{equation}
        \phi_{hard}(g_k(x_i), g_k(x_j)) = M g_k(x_i) + (1-M) g_k(x_j)
        \label{eq:Hard patch up}
    \end{equation}
    Where M represents a binary mask. On the other hand, Soft PatchUp's operation at layer k is defined as in \eqref{eq:Soft patch up}:
    \begin{equation}
        \phi_{soft}(g_k(x_i), g_k(x_j)) = M g_k(x_i) + Mix_\lambda[((1-M) g_k(x_i)), ((1-M) g_k(x_j))]
        \label{eq:Soft patch up}
    \end{equation}
    The loss function 
    \begin{equation}
      \text{loss}_{\text{patchup}} = \mathbb{E}_{(x_i, y_i) \sim P} \mathbb{E}_{(x_j, y_j) \sim P} \mathbb{E}_{\lambda \sim \text{Beta}(\alpha, \alpha)} \mathbb{E}_{k \sim S} \text{Mix}_{\text{pu}}\left[\left(f_k(\phi_k), y_i\right), \left(f_k(\phi_k), Y\right)\right]
      \label{eq:patchup loss}
    \end{equation}

    where $\text{pu}$ is the fraction of unaltered features in the unaltered feature map, $S$ is the set of layers where PatchUp is randomly applied, and $\phi$ refers to $\phi_{\text{hard}}$ for Hard PatchUp and $\phi_{\text{soft}}$ for Soft PatchUp. $Y$ denotes the target associated with the altered features. In the case of Hard PatchUp, $Y = y_j$, and for Soft PatchUp, $Y = \text{Mix}_\lambda(y_i, y_j)$. The reweighted target $W(y_i, y_j)$ is computed based on the interpolation strategy of $y_i$ and $y_j$. The $W$ definitions for Hard PatchUp and Soft PatchUp are given as\cite{faramarzi2020patchup}:
    
    \begin{align}
      W_{\text{hard}}(y_i, y_j) &= \text{Mix}_{\text{pu}}(y_i, y_j) \\
      W_{\text{soft}}(y_i, y_j) &= \text{Mix}_{\text{pu}}(y_i, \text{Mix}_\lambda(y_i, y_j))\\
      \text{Mix}_\lambda(a, b) &= \lambda \cdot a + (1 - \lambda) \cdot b
    \end{align}
    With $\lambda \in [0,1]$ as the mixing coefficient, which is sampled from a Beta distribution. The sample process of patchup has been illustrated in Figure \ref{fig 3.3:Patch up}.In my implementation, the soft patchup is performed on the hidden layer pseudo sequence part. For hyper-parameter selection, patchup prob, $\gamma$, and block size equals to 1.0, 0.75,and 1.
    
    \section{CC(Cam-Center) Loss}
    Inspired by the grad-CAM\cite{selvaraju2017grad}, I proposed a novel loss criterion named as CC loss(cam-center loss). My loss function is shown below in \eqref{eq:Total loss}
    \begin{equation}
      \text{loss} = \text{loss}_{\text{clip\_BCE}} + \text{loss}_{\text{patchup\_soft}} + \text{loss}_{\text{CC}}
      \label{eq:Total loss}
    \end{equation}
    the clip BCE loss is defined as \eqref{eq:loss_clip}, and patchup loss can refer to \eqref{eq:patchup loss}
    \begin{equation}
      \text{loss}_{\text{clip\_BCE}} = -\frac{1}{N} \sum_{i=1}^{N} \left[y_i \cdot \text{clip}(\log(\hat{y}_i)) + (1 - y_i) \cdot \text{clip}(\log(1 - \hat{y}_i))\right]
      \label{eq:loss_clip}
    \end{equation}
    where the \textit{clip} function is defined as \eqref{eq:clip}:
    \begin{equation}
      \text{clip}(x) = 
      \begin{cases}
        x & \text{if } 0.25 \leq \exp(x) \leq 0.75 \\
        0 & \text{otherwise}
      \end{cases}
      \label{eq:clip}
    \end{equation}
    \begin{algorithm}
    \caption{Training process with CC-loss}
    \label{alg:example}
    \begin{algorithmic}[1]
    \REQUIRE Input parameters $v_i$
    \ENSURE Output results  $out_i$
    \STATE \textbf{Initialization:}the parameters of backbone $\theta$; The all-zero binary-class instruction map $m_0$ and $m_1$
    \STATE \textbf{Optimization:}
    \FOR {each training iteration}
        \IF {iteration<100}
            \STATE clear the map $m_0$ and $m_1$
        \ENDIF
        \STATE Forward the input through the network and hook the feature map $f_i$ of the CCT last fully connection layer output.
        \STATE Take the average feature of each class to build $m_0$ and $m_1$
        \STATE Calculate $\text{loss}_{\text{clip\_BCE}}$ then Backward once to hook the gradient $G$ of CCT last fully connection layer.
        \STATE Normalize the $G$ between 0-1 
        \STATE Calculate the CC-loss
        \begin{equation}
        \text{CC-loss} = \frac{1}{N}\sum_{i=1}^{N}\left[ y_i \cdot |f_i^+ - m_1| \odot G + (1 - y_i) \cdot |f_i^- - m_0| \odot G \right]
        \end{equation}
        where:
        \begin{itemize}
        \item $N$: total number of samples
        \item $y_i$: label of the $i$-th sample (1 for positive class, 0 for negative class)
        \item $f_i^+$: feature map of the $i$-th positive class sample
        \item $f_i^-$: feature map of the $i$-th negative class sample
        \item $\odot$: element-wise multiplication (Hadamard product)
        \end{itemize}

        \STATE Clear the gradient then backward again with the total loss
        \STATE Update the $\theta \leftarrow \theta - \eta \cdot \nabla_\theta \text{CC-loss}$
    \ENDFOR
    \STATE Return optimal parameters $\theta*$
    \end{algorithmic}
    \end{algorithm}
    \chapter{Experiment Result}
    \section{Dataset And Preprocessing}
    Before explain my motivation, firstly I need to introduce the dataset and related pre-processing. 
    the initial 24-hour ICU records utilized for predicting in-hospital mortality (Figure 1). The raw data were discretized into 1-hour intervals \cite{harutyunyan2019multitask}, and 812 variables, including those mentioned in \cite{nie2021mortality} and \cite{gao2020stagenet}, were used without manual selection .
    
    Discretized data gaps were addressed using forward imputation, and when previous data were unavailable, average values or a '-1' placeholder were employed. Categorical variables underwent one-hot encoding, with discrete values transformed into binary bit groups. Data were pre-normalized based on the training set's distribution after discretization and imputation.\cite{Guo2022}
    
    A total of 721 samples were extracted from the MIMIC-III database,with 199 positive (death)class and 522 negative (Live)class, resulting in a relatively small and imbalance dataset. To enhance robustness, 10-fold cross-validation was conducted on the dataset.
    
    \section{Statistical Analysis}
    The Area Under the Receiver Operating Characteristic curve (AUROC) and confusion matrix are both widely used performance evaluation metrics for binary classification problems. The AUROC quantifies the classifier's ability to discriminate between positive and negative classes, whereas the confusion matrix provides a more detailed analysis of true positives (TP), false positives (FP), true negatives (TN), and false negatives (FN).

    The confusion matrix can be represented as \ref{tab:confusion_matrix}
    \begin{table}[h!]
    \centering
    \begin{tabular}{c|c|c|c}
     & & \multicolumn{2}{c}{\textbf{Predicted}} \\\cline{1-4}
     & & \textbf{Positive} & \textbf{Negative} \\ \cline{1-4}
    \multirow{2}{*}{\textbf{Actual}} & \textbf{Positive} & TP & FN \\
     & \textbf{Negative} & FP & TN \\ \cline{1-4}
    \end{tabular}
    \caption{Confusion Matrix}
    \label{tab:confusion_matrix}
    \end{table}
    
    From the confusion matrix, we can derive various performance metrics such as precision, recall, and F1-score. However, these metrics are threshold-dependent, which means they depend on the specific threshold chosen to classify an instance as positive or negative.
    
    On the other hand, the AUROC provides a threshold-independent measure of classification performance. The ROC curve is generated by plotting the true positive rate (TPR) against the false positive rate (FPR) at various threshold settings. Mathematically, TPR and FPR are defined as follows:
    
    \begin{equation}
    \text{Sensitivity} =\text{TPR}= \frac{\text{TP}}{\text{TP} + \text{FN}}
    \end{equation}
    \begin{equation}
    \text{Specificity} =\text{fPR}= \frac{\text{TN}}{\text{TN} + \text{FP}}
    \end{equation}
    \begin{equation}
    \text{Accuracy} = \frac{\text{TP}+\text{TN}}{\text{TP}+\text{FP}+\text{TN} + \text{FN}}
    \end{equation}

    The area under the ROC curve (AUROC) provides a single value that represents the classifier's performance across all possible threshold values. An AUROC of 0.5 indicates random classification, while an AUROC of 1.0 signifies perfect classification.
    The AUROC provides a threshold-independent evaluation metric, while the confusion matrix offers a more detailed, threshold-dependent analysis of classifier performance.
    \section{Motivation And Experiment}
    \subsection{Selection Of Transformer}
        At the beginning I applied the vanilla transformer \cite{vaswani2017attention},tab-transformer \cite{huang2020tabtransformer} and FT-transformer \cite{gorishniy2021revisiting}. Experiment shown below in Table \ref{tbl:transformer comparison}
        
        \begin{table}[!htbp]

\caption{Transformer Performance Comparison}
\label{tbl:transformer comparison}
\begin{tabular}{lcccc}
\toprule
Implement & Sensitivity & Specificity & Accuracy & AUROC \\
\midrule
Random forest (baseline) & \makecell{$0.6973$ \\ $\pm 0.0883$} & \makecell{$0.9075$ \\ $\pm 0.0375$} & \makecell{$0.8502$ \\ $\pm 0.0357$} & \makecell{$0.8024$ \\ $\pm 0.0451$} \\
ResNet+StageNet (baseline) & \makecell{$0.7833$ \\ $\pm 0.1123$} & \makecell{$\Large\mathbf{0.8648}$ \\ $\pm 0.0662$} & \makecell{$\Large\mathbf{0.8444}$ \\ $\pm 0.0339$} & \makecell{$0.9166$ \\ $\pm 0.0132$} \\
StageNet+vanilla transformer & \makecell{$0.7000$ \\ $ \pm 0.0235$} & \makecell{$0.7870$ \\ $\pm 0.0083$} & \makecell{$0.7652$ \\ $\pm 0.0034$} & \makecell{$0.7955$ \\ $\pm 0.0009$} \\
Tab transformer & \makecell{$0.7944$ \\ $\pm 0.0074$} & \makecell{$0.8018$ \\ $\pm 0.0071$} & \makecell{$0.8000$ \\ $\pm 0.0025$} & \makecell{$0.8838$ \\ $\pm 0.0001$} \\
FT transformer &  \makecell{$0.7999$ \\ $\pm 0.0026$} & \makecell{$0.7888$ \\ $\pm 0.0023$} & \makecell{$0.7917$ \\ $ \pm 0.0008$} & \makecell{$0.8647$ \\ $ \pm 0.0008$} \\

CCT(CIFAR-100 pretrained) & \makecell{$\Large\mathbf{0.8500}$ \\ $\pm 0.0204$} & \makecell{$0.7815$ \\ $\pm 0.0102$} & \makecell{$0.7986$ \\ $\pm 0.0023$} & \makecell{$\Large\mathbf{0.9189}$ \\ $\pm 0.0001$} \\
\bottomrule
\end{tabular}
\end{table}

        The experimental results were perplexing, as it is generally expected that transformers should outperform LSTMs in terms of performance. However, the network's performance appeared to be worse. To understand this phenomenon, I consulted a research article \cite{matsoukas2021time}that explained and demonstrated that transformers, compared to CNNs and other neural networks, require larger amounts of data to achieve comparable or even superior performance. In the medical domain, data confidentiality is crucial, and many datasets do not point to the same patient, with variables not always being the same. Consequently, obtaining more data is virtually impossible. Thus, is there a method that enables transformers to achieve good performance on small datasets?

        According to the article\cite{matsoukas2021time}, there are two possible approaches: self-supervised learning and transfer learning. The self-supervised learning mentioned in the article involves data augmentation using adversarial tasks on images, which is not applicable to clinical data. As for transfer learning, its value for pretraining on tabular data is limited, mainly due to the small dataset sizes and the significant difference between the task domains (multivariate time series classification).
        
        Considering that my data is also two-dimensional, I wondered why I should not utilize ViT. Based on this thought, the paper of CCT\cite{hassani2021cct} proposed a ViT-based model with good performance on small datasets and pretrained weights available for ImageNet-1k. This approach has the potential to enable transformers to perform well on smaller medical datasets while maintaining the advantages of transformer-based models.

    \subsection{With Or Without Sequence}
        Initially, in order to preserve the original time series, we replaced the LSTM cell with the CCT, aiming to maintain the temporal dependencies within the data. However, this approach resulted in the loss of interconnections between the cells, transforming the role of CCT into another feature extractor. Consequently, we attempted to replace the entire Stage-aware LSTM Module with CCT. The trade-off of this approach was that the obtained sequences became pseudo-sequences, and the experimental results are presented as follows in table \ref{tbl:sequence comparison}.
        \begin{table}[!htbp]

        \caption{Sequence Performance Comparison }
        \label{tbl:sequence comparison}
        \begin{tabular}{lcccc}
        \toprule
        Implement & Sensitivity & Specificity & Accuracy & AUROC \\
        \midrule
        ResNet+StageNet (baseline) & \makecell{$0.7833$ \\ $\pm 0.1123$} & \makecell{$0.8648$ \\ $\pm 0.0662$} & \makecell{$0.8444$ \\ $\pm 0.0339$} & \makecell{$0.9166$ \\ $\pm 0.0132$} \\
        CCT (cifar-100)& \makecell{$\Large\mathbf{0.8500}$ \\ $\pm 0.0204$} & \makecell{$0.7815$ \\ $\pm 0.0102$} & \makecell{$0.7986$ \\ $\pm 0.0023$} & \makecell{$0.9189$ \\ $\pm 0.0001$} \\
        Pseudo-seq CCT(freeze tokenizer) & \makecell{$0.8055$ \\ $ \pm 0.0013$} & \makecell{$\Large\mathbf{0.8777}$ \\ $\pm 0.0006$} & \makecell{$\Large\mathbf{0.8597}$ \\ $\pm 0.0004$} & \makecell{$\Large\mathbf{0.9325}$ \\ $\pm 0.0000$} \\
        (not freeze) & \makecell{$0.7944$ \\ $ \pm 0.0154$} & \makecell{$\Large\mathbf{0.8407}$ \\ $\pm 0.0038$} & \makecell{$\Large\mathbf{0.8291}$ \\ $\pm 0.0004$} & \makecell{$\Large\mathbf{0.8898}$ \\ $\pm 0.0000$} \\
        \bottomrule
        \end{tabular}
        \end{table}
        
        As demonstrated by the experimental results, we observe a substantial improvement in the performance of the pseudo-sequence compared to prior approaches. It is important to note that the tokenizer of the pre-trained model needs to be frozen; not doing so may result in a decline in model performance. A possible reason for this is the difference between the hidden layer images of tabular data and the original images. The corresponding experimental results are shown in the table.

        This enhanced performance establishes a robust foundation for carrying out subsequent experiments and investigations. In future endeavors, we will continue to build upon this foundation, striving to further optimize the model to achieve superior outcomes, while taking into consideration the need to freeze the tokenizer of the pre-trained model.
    \subsection{Mixup And CC-loss}
        Considering that our dataset is relatively small and imbalanced, we hypothesize that data augmentation techniques could potentially enhance the performance of our model. However, conventional data augmentation methods, such as flipping and cropping, are not applicable to tabular data. As a result, we opt for the mixup\cite{zhang2017mixup} technique. Due to the presence of missing data in our dataset, directly applying mixup to the input yields suboptimal results. Therefore, we focus on implementing mixup for intermediate hidden layers.

        There are two primary methods for hidden layer mixup: Manifold Mixup\cite{verma2019manifold} and Patch Up\cite{faramarzi2020patchup}. It is important to note that in Manifold Mixup, the authors mentioned that the deeper the layer where mixup is applied, the lower the training set loss. We conducted experiments with both Manifold Mixup and Patch Up to evaluate their impact on our model's performance, with the results presented below:

        \begin{table}[!htbp]
    \caption{Augmentation and metric learning Comparison}
    \label{tbl:mixup and CC loss comparison}
    \begin{tabular}{lcccc}
        \toprule
        Implement & Sensitivity & Specificity & Accuracy & AUROC \\
        \midrule
        ResNet+StageNet (baseline) & \makecell{$0.7833$ \\ $\pm 0.1123$} & \makecell{$0.8648$ \\ $\pm 0.0662$} & \makecell{$0.8444$ \\ $\pm 0.0339$} & \makecell{$0.9166$ \\ $\pm 0.0132$} \\
        Pseudo-sequence CCT (imagenet) & \makecell{$\Large\mathbf{0.8055}$ \\ $\pm 0.0013$} & \makecell{$0.8777$ \\ $\pm 0.0006$} & \makecell{$0.8597$ \\ $\pm 0.0004$} & \makecell{$0.9325$ \\ $\pm 0.0000$} \\
        (manifold mix-up) & \makecell{$0.6500$ \\ $\pm 0.0345$} & \makecell{$0.9277$ \\ $\pm 0.0014$} & \makecell{$0.8583$ \\ $\pm 0.0007$} & \makecell{$0.9317$ \\ $\pm 0.0000$} \\
        (Patch-up hard) & \makecell{$0.6999$ \\ $\pm 0.0125$} & \makecell{$0.9296$ \\ $\pm 0.0028$} & \makecell{$0.8722$ \\ $\pm 0.0006$} & \makecell{$0.9299$ \\ $\pm 0.0001$} \\
        (Patch-up soft) & \makecell{$0.7944$ \\ $\pm 0.0099$} & \makecell{$0.9055$ \\ $\pm 0.0037$} & \makecell{$0.8777$ \\ $\pm 0.0005$} & \makecell{$0.9354$ \\ $\pm 0.0000$} \\
        CC-loss & \makecell{$0.5722$ \\ $\pm 0.0075$} & \makecell{$\Large\mathbf{0.9648}$ \\ $\pm 0.0005$} & \makecell{$0.8666$ \\ $\pm 0.0001$} & \makecell{$0.9345$ \\ $\pm 0.0000$} \\
        CC-loss+patch up-soft & \makecell{$0.7611$ \\ $\pm 0.0062$} & \makecell{$0.9166$ \\ $\pm 0.0026$} & \makecell{$\Large\mathbf{0.8777}$ \\ $\pm 0.0004$} & \makecell{$\Large\mathbf{0.9415}$ \\ $\pm 0.0000$} \\
        No-Stage-adaptive (Ablation Study)& \makecell{$0.8166$ \\ $\pm 0.0111$} & \makecell{$0.8722$ \\ $\pm 0.0039$} & \makecell{$0.8583$ \\ $\pm 0.0004$} & \makecell{$0.9269$ \\ $\pm 0.0001$} \\
        \bottomrule
    \end{tabular}
    \end{table}
    The experimental results clearly demonstrate that among the three mentioned mixup methods, only PatchUp Soft leads to an improvement in the network's performance.

    Furthermore, it is reasonable to assume that two classes each have distinct centers. However, due to differences in individual data and issues with missing data, there will always be samples closer to the other class's center. To address this issue, the proposed idea is to train the model to approach the correct center while moving away from incorrect samples. Simultaneously, attention mechanisms, specifically in the form of Class Activation Mapping (CAM), are introduced for weighting purposes. This approach emphasizes important features that contribute to the accurate classification of samples.
    
    Based on these considerations, the aforementioned CamCenterLoss (CC-loss) is proposed. While the primary objective of the original loss function is to separate two classes using a high-dimensional plane, the goal of CC-loss is to utilize attention in the hidden layers and reduce the inter-class distance in an element-wise manner. This, in turn, enhances the performance of classification tasks. 
    
    According to the experimental results, we observed that the CC-loss method enhanced the model performance by approximately 0.2$\%$, while the Patch Up soft method improved it by around 0.3$\%$. Remarkably, when employing both methods simultaneously, the model performance experienced a significant boost of 0.9$\%$.

    Furthermore, in the ablation study, we aimed to verify whether the Stage-adaptive Convolutional Module could be replaced by a fully connected layer. The experimental results, presented in the accompanying table, demonstrate a decline in model performance when this structure is removed. This evidence substantiates the importance and effectiveness of the Stage-adaptive Convolutional Module in our model.

\chapter{Conclusion And Future Work}
\section{Conclusion}

In this study, we aimed to develop an effective method for predicting in-hospital mortality using a relatively small and imbalanced dataset extracted from the MIMIC-III database. By employing a combination of transformer-based models and data augmentation techniques, we demonstrated the potential of our approach in achieving improved performance on this challenging task. Our key contributions can be summarized as follows:

We investigated the use of the CCT model, a ViT-based architecture, which exhibited promising results on small datasets, despite transformers' general dependency on larger datasets for superior performance.

We explored the impact of utilizing pseudo-sequences instead of real sequences in our model and the importance of freezing the tokenizer of the pretrained model. Not freezing the tokenizer led to a decline in the model's performance, likely due to the differences between the hidden layer images of tabular data and those of the original images. Our findings suggest that transformers pretrained on images can also perform well on pseudo-time sequences.

We assessed the effectiveness of mixup techniques, specifically Manifold Mixup and Patch Up, for data augmentation in tabular data. Our findings revealed that PatchUp Soft is the most effective mixup method, leading to an enhancement in our model's performance.

We proposed the novel CC-loss method, a metric learning technique that leverages attention mechanisms to reduce inter-class distance in an element-wise manner, consequently improving the performance of classification tasks.

Our ablation study confirmed the importance of the Stage-adaptive Convolutional Module in our model, as its removal resulted in a decline in performance.

In conclusion, our proposed approach demonstrates that combining transformer-based models with data augmentation techniques, such as Patch Up Soft and the metric learning method CC-loss, can lead to substantial improvements in predicting in-hospital mortality using small and imbalanced datasets. This study not only contributes to the field of mortality prediction in the medical domain but also highlights the potential of transformer-based models for small datasets and pseudo-time sequences, opening up avenues for future research and practical applications in other fields with similar data constraints.

\section{Future Work}

Our current approach demonstrates promising results in predicting in-hospital mortality using small and imbalanced datasets. However, there are several avenues for improvement and exploration to further enhance the model's performance:

\textbf{1. Incorporate multimodal data sources:} Integrating more diverse and comprehensive data sources, such as electronic health records, lab results, or imaging data, could potentially improve the model's performance by providing a richer representation of patients' medical conditions and further aid in mortality prediction.

\textbf{2. Design a new CCT to maintain sequences while using pretrained tokenizer:} One potential direction is to explore the development of novel CCT architectures that can preserve the sequence information while utilizing the pretrained tokenizer. This could potentially lead to better performance in handling time series data and improve the overall predictive capability of the model.

\textbf{3. Test different hyperparameters or methods based on PatchUp:} Investigate the effects of varying hyperparameters and methods within the PatchUp framework, which could lead to further improvements in the model's performance. By experimenting with different combinations and strategies, it may be possible to identify more effective data augmentation techniques tailored for our specific task and dataset.

\textbf{4. Address class imbalance more effectively:} Explore alternative techniques for handling class imbalance in the dataset, such as re-sampling, cost-sensitive learning, or ensemble methods, to improve the model's performance, particularly in terms of sensitivity and specificity.

By addressing these potential areas of improvement and exploration, future work could lead to even better performance in mortality prediction and contribute to the development of more effective and reliable tools for healthcare professionals, ultimately improving patient care and outcomes.

%
%
%
%
%
%
%
	
\balance
\bibliographystyle{unsrt} 
\bibliography{bibliography.bib}

\begin{thebibliography}{10}

\bibitem{feigin2003stroke}
Valery~L Feigin, Carlene~MM Lawes, Derrick~A Bennett, and Craig~S Anderson.
\newblock Stroke epidemiology: a review of population-based studies of
  incidence, prevalence, and case-fatality in the late 20th century.
\newblock {\em The lancet neurology}, 2(1):43--53, 2003.

\bibitem{cordonnier2018intracerebral}
Charlotte Cordonnier, Andrew Demchuk, Wendy Ziai, and Craig~S Anderson.
\newblock Intracerebral haemorrhage: current approaches to acute management.
\newblock {\em The Lancet}, 392(10154):1257--1268, 2018.

\bibitem{van2010incidence}
Charlotte~JJ Van~Asch, Merel~JA Luitse, Gabri{\"e}l~JE Rinkel, Ingeborg van~der
  Tweel, Ale Algra, and Catharina~JM Klijn.
\newblock Incidence, case fatality, and functional outcome of intracerebral
  haemorrhage over time, according to age, sex, and ethnic origin: a systematic
  review and meta-analysis.
\newblock {\em The Lancet Neurology}, 9(2):167--176, 2010.

\bibitem{carolei1997high}
Antonio Carolei, Carmine Marini, Mario Di~Napoli, Giacinto Di~Gianfilippo,
  Paola Santalucia, Massimo Baldassarre, Giorgio De~Matteis, and Ferdinando
  Di~Orio.
\newblock High stroke incidence in the prospective community-based l’aquila
  registry (1994--1998) first year’s results.
\newblock {\em Stroke}, 28(12):2500--2506, 1997.

\bibitem{monteiro2018using}
Miguel Monteiro, Ana~Catarina Fonseca, Ana~Teresa Freitas, Teresa~Pinho e~Melo,
  Alexandre~P Francisco, Jose~M Ferro, and Arlindo~L Oliveira.
\newblock Using machine learning to improve the prediction of functional
  outcome in ischemic stroke patients.
\newblock {\em IEEE/ACM transactions on computational biology and
  bioinformatics}, 15(6):1953--1959, 2018.

\bibitem{nie2021mortality}
Ximing Nie, Yuan Cai, Jingyi Liu, Xiran Liu, Jiahui Zhao, Zhonghua Yang, Miao
  Wen, and Liping Liu.
\newblock Mortality prediction in cerebral hemorrhage patients using machine
  learning algorithms in intensive care units.
\newblock {\em Frontiers in Neurology}, 11:610531, 2021.

\bibitem{calasan2019}
Martin {\'C}alasan, Danilo Muji{\v{c}}i{\'c}, Vesna Rube{\v{z}}i{\'c}, and
  Milovan Radulovi{\'c}.
\newblock Estimation of equivalent circuit parameters of single-phase
  transformer by using chaotic optimization approach.
\newblock {\em Energies}, 12(9):1697, 2019.

\bibitem{strain2019}
Nathan~N. Strain, Jingjing Sun, Xingxuan Huang, Daniel~J. Costinett, and
  Leon~M. Tolbert.
\newblock Zvs analysis of a gan-based series-parallel dual transformer llc
  resonant converter.
\newblock In {\em 2019 IEEE 7th Workshop on Wide Bandgap Power Devices and
  Applications (WiPDA)}, pages 398--404, 2019.

\bibitem{palomino2020}
Alejandro Palomino and Masood Parvania.
\newblock Data-driven risk analysis of joint electric vehicle and solar
  operation in distribution networks.
\newblock {\em IEEE Open Access Journal of Power and Energy}, 7:141--150, 2020.

\bibitem{li2021}
Bo~Li, Ruoming Pang, Tara~N Sainath, Anmol Gulati, Yu~Zhang, James Qin, Parisa
  Haghani, W~Ronny Huang, Min Ma, and Junwen Bai.
\newblock Scaling end-to-end models for large-scale multilingual asr.
\newblock In {\em 2021 IEEE Automatic Speech Recognition and Understanding
  Workshop (ASRU)}, pages 1011--1018. IEEE, 2021.

\bibitem{park2021}
Sangjoon Park, Gwanghyun Kim, Yujin Oh, Joon~Beom Seo, Sang~Min Lee, Jin~Hwan
  Kim, Sungjun Moon, Jae-Kwang Lim, and Jong~Chul Ye.
\newblock Vision transformer for covid-19 cxr diagnosis using chest x-ray
  feature corpus, 2021.

\bibitem{gao2021}
Xiaohong Gao, Yu~Qian, and Alice Gao.
\newblock Covid-vit: Classification of covid-19 from ct chest images based on
  vision transformer models, 2021.

\bibitem{pan2021}
Jiayi Pan, Heye Zhang, Weifei Wu, Zhifan Gao, and Weiwen Wu.
\newblock Multi-domain integrative swin transformer network for sparse-view
  tomographic reconstruction.
\newblock {\em Patterns}, 3(6):100498, 2022.

\bibitem{xie2022}
Huiqiang Xie, Zhijin Qin, Xiaoming Tao, and Khaled~B Letaief.
\newblock Task-oriented multi-user semantic communications.
\newblock {\em IEEE Journal on Selected Areas in Communications},
  40(9):2584--2597, 2022.

\bibitem{lin2022}
Nan Lin, Shen Dong, and Yihong Liu.
\newblock Transformer fault diagnosis method based on ifcm-dnn adjudication
  network.
\newblock In {\em International Conference on Intelligent Traffic Systems and
  Smart City (ITSSC 2021)}, volume 12165, pages 640--652. SPIE, 2022.

\bibitem{keitoue2018}
Samir Keitoue, Ivan Murat, Bo{\v{z}}idar Filipovi{\'c}-Gr{\v{c}}i{\'c}, Alan
  {\v{Z}}upan, Ivana Damjanovi{\'c}, and Ivica Pavi{\'c}.
\newblock Lightning caused overvoltages on power transformers recorded by
  on-line transient overvoltage monitoring system.
\newblock {\em Journal of Energy: Energija}, 67(2):0--0, 2018.

\bibitem{harutyunyan2019multitask}
Hrayr Harutyunyan, Hrant Khachatrian, David~C Kale, Greg Ver~Steeg, and Aram
  Galstyan.
\newblock Multitask learning and benchmarking with clinical time series data.
\newblock {\em Scientific data}, 6(1):96, 2019.

\bibitem{gao2020stagenet}
Junyi Gao, Cao Xiao, Yasha Wang, Wen Tang, Lucas~M. Glass, and Jimeng Sun.
\newblock Stagenet: Stage-aware neural networks for health risk prediction.
\newblock In Yennun Huang, Irwin King, Tie{-}Yan Liu, and Maarten van Steen,
  editors, {\em {WWW} '20: The Web Conference 2020, Taipei, Taiwan, April
  20-24, 2020}, pages 530--540. {ACM} / {IW3C2}, 2020.

\bibitem{Guo2022}
JingYuan Guo.
\newblock Deep learning for analysing brain medical data.
\newblock Master's thesis, National University of Singapore, 2022.

\bibitem{hassani2021cct}
Ali Hassani, Steven Walton, Nikhil Shah, Abulikemu Abuduweili, Jiachen Li, and
  Humphrey Shi.
\newblock Escaping the big data paradigm with compact transformers.
\newblock {\em ArXiv preprint}, abs/2104.05704, 2021.

\bibitem{dosovitskiy2020image}
Alexey Dosovitskiy, Lucas Beyer, Alexander Kolesnikov, Dirk Weissenborn,
  Xiaohua Zhai, Thomas Unterthiner, Mostafa Dehghani, Matthias Minderer, Georg
  Heigold, Sylvain Gelly, Jakob Uszkoreit, and Neil Houlsby.
\newblock An image is worth 16x16 words: Transformers for image recognition at
  scale.
\newblock In {\em 9th International Conference on Learning Representations,
  {ICLR} 2021, Virtual Event, Austria, May 3-7, 2021}. OpenReview.net, 2021.

\bibitem{faramarzi2020patchup}
Mojtaba Faramarzi, Mohammad Amini, Akilesh Badrinaaraayanan, Vikas Verma, and
  Sarath Chandar.
\newblock Patchup: A regularization technique for convolutional neural
  networks.
\newblock {\em ArXiv preprint}, abs/2006.07794, 2020.

\bibitem{selvaraju2017grad}
Ramprasaath~R. Selvaraju, Michael Cogswell, Abhishek Das, Ramakrishna Vedantam,
  Devi Parikh, and Dhruv Batra.
\newblock Grad-cam: Visual explanations from deep networks via gradient-based
  localization.
\newblock In {\em {IEEE} International Conference on Computer Vision, {ICCV}
  2017, Venice, Italy, October 22-29, 2017}, pages 618--626. {IEEE} Computer
  Society, 2017.

\bibitem{vaswani2017attention}
Ashish Vaswani, Noam Shazeer, Niki Parmar, Jakob Uszkoreit, Llion Jones,
  Aidan~N. Gomez, Lukasz Kaiser, and Illia Polosukhin.
\newblock Attention is all you need.
\newblock In Isabelle Guyon, Ulrike von Luxburg, Samy Bengio, Hanna~M. Wallach,
  Rob Fergus, S.~V.~N. Vishwanathan, and Roman Garnett, editors, {\em Advances
  in Neural Information Processing Systems 30: Annual Conference on Neural
  Information Processing Systems 2017, December 4-9, 2017, Long Beach, CA,
  {USA}}, pages 5998--6008, 2017.

\bibitem{huang2020tabtransformer}
Xin Huang, Ashish Khetan, Milan Cvitkovic, and Zohar Karnin.
\newblock Tabtransformer: Tabular data modeling using contextual embeddings.
\newblock {\em ArXiv preprint}, abs/2012.06678, 2020.

\bibitem{gorishniy2021revisiting}
Yury Gorishniy, Ivan Rubachev, Valentin Khrulkov, and Artem Babenko.
\newblock Revisiting deep learning models for tabular data.
\newblock In Marc'Aurelio Ranzato, Alina Beygelzimer, Yann~N. Dauphin, Percy
  Liang, and Jennifer~Wortman Vaughan, editors, {\em Advances in Neural
  Information Processing Systems 34: Annual Conference on Neural Information
  Processing Systems 2021, NeurIPS 2021, December 6-14, 2021, virtual}, pages
  18932--18943, 2021.

\bibitem{matsoukas2021time}
Christos Matsoukas, Johan~Fredin Haslum, Magnus S{\"o}derberg, and Kevin Smith.
\newblock Is it time to replace cnns with transformers for medical images?
\newblock {\em ArXiv preprint}, abs/2108.09038, 2021.

\bibitem{zhang2017mixup}
Hongyi Zhang, Moustapha Ciss{\'{e}}, Yann~N. Dauphin, and David Lopez{-}Paz.
\newblock mixup: Beyond empirical risk minimization.
\newblock In {\em 6th International Conference on Learning Representations,
  {ICLR} 2018, Vancouver, BC, Canada, April 30 - May 3, 2018, Conference Track
  Proceedings}. OpenReview.net, 2018.

\bibitem{verma2019manifold}
Vikas Verma, Alex Lamb, Christopher Beckham, Amir Najafi, Ioannis Mitliagkas,
  David Lopez-Paz, and Yoshua Bengio.
\newblock Manifold mixup: Better representations by interpolating hidden
  states.
\newblock In {\em International conference on machine learning}, pages
  6438--6447. PMLR, 2019.

\end{thebibliography}

\end{document}